\documentclass[9pt,conference]{IEEEtran}
\IEEEoverridecommandlockouts
\usepackage{cite}
\usepackage{amsmath,amssymb,amsfonts}
\usepackage{algorithmic}
\usepackage{graphicx}
\usepackage{textcomp}
\usepackage{xcolor}
\usepackage{url}
\usepackage{subfigure}
\usepackage{mathtools}
\usepackage{amsthm}
\usepackage{bm}
\usepackage{multicol,multirow,adjustbox,algorithm}
\def\BibTeX{{\rm B\kern-.05em{\sc i\kern-.025em b}\kern-.08em
    T\kern-.1667em\lower.7ex\hbox{E}\kern-.125emX}}
\begin{document}

\title{VALL-T: Decoder-Only Generative Transducer for Robust and Decoding-Controllable Text-to-Speech
% {\footnotesize \textsuperscript{*}Note: Sub-titles are not captured for https://ieeexplore.ieee.org  and
% should not be used}
}

% \title{VALL-T: Decoder-Only Generative Transducer for Robust and Decoding-Controllable Text-to-Speech\\
% % {\footnotesize \textsuperscript{*}Note: Sub-titles are not captured for https://ieeexplore.ieee.org  and
% % should not be used}
% % \thanks{Identify applicable funding agency here. If none, delete this.}
% }

\author{\IEEEauthorblockN{Chenpeng Du$^1$\textsuperscript{*}\thanks{\textsuperscript{*}Main contributors.}, Yiwei Guo$^1$\textsuperscript{*}, Hankun Wang$^1$\textsuperscript{*}, Yifan Yang$^1$, Zhikang Niu$^1$,\\Shuai Wang$^2$, Hui Zhang$^3$, Xie Chen$^1$, Kai Yu$^1$\textsuperscript{\textdagger}\thanks{\textsuperscript{*}Kai Yu is the corresponding author.}}
\IEEEauthorblockA{\textit{$^1$MoE Key Lab of Artificial Intelligence, AI Institute ; X-LANCE Lab, Shanghai Jiao Tong University, Shanghai, China}}
\IEEEauthorblockA{\textit{$^{2}$Shenzhen Research Institute of Big Data, Shenzhen, China}}
\IEEEauthorblockA{\textit{$^{3}$AISpeech, Beijing, China}}
% \and
% \IEEEauthorblockN{2\textsuperscript{nd} Given Name Surname}
% \IEEEauthorblockA{\textit{dept. name of organization (of Aff.)} \\
% \textit{name of organization (of Aff.)}\\
% City, Country \\
% email address or ORCID}
% \and
% \IEEEauthorblockN{3\textsuperscript{rd} Given Name Surname}
% \IEEEauthorblockA{\textit{dept. name of organization (of Aff.)} \\
% \textit{name of organization (of Aff.)}\\
% City, Country \\
% email address or ORCID}
% \and
% \IEEEauthorblockN{4\textsuperscript{th} Given Name Surname}
% \IEEEauthorblockA{\textit{dept. name of organization (of Aff.)} \\
% \textit{name of organization (of Aff.)}\\
% City, Country \\
% email address or ORCID}
% \and
% \IEEEauthorblockN{5\textsuperscript{th} Given Name Surname}
% \IEEEauthorblockA{\textit{dept. name of organization (of Aff.)} \\
% \textit{name of organization (of Aff.)}\\
% City, Country \\
% email address or ORCID}
% \and
% \IEEEauthorblockN{6\textsuperscript{th} Given Name Surname}
% \IEEEauthorblockA{\textit{dept. name of organization (of Aff.)} \\
% \textit{name of organization (of Aff.)}\\
% City, Country \\
% email address or ORCID}
}

\maketitle

\begin{abstract}
Recent TTS models with decoder-only Transformer architecture, such as SPEAR-TTS and VALL-E, achieve impressive naturalness and demonstrate the ability for zero-shot adaptation given a speech prompt. However, such decoder-only TTS models lack monotonic alignment constraints, sometimes leading to hallucination issues such as mispronunciation, word skipping and repeating. To address this limitation, we propose VALL-T, a generative Transducer model that introduces shifting relative position embeddings for input phoneme sequence, explicitly indicating the monotonic generation process while maintaining the architecture of decoder-only Transformer. Consequently, VALL-T retains the capability of prompt-based zero-shot adaptation and demonstrates better robustness against hallucinations with a relative reduction of 28.3\% in the word error rate. 
% Furthermore, the controllability of alignment in VALL-T during decoding facilitates the use of untranscribed speech prompts within in-context learning framework.
The audio samples are available at \texttt{\url{https://cpdu.github.io/vallt}}.
\end{abstract}

\begin{IEEEkeywords}
transducer, text-to-speech, decoder-only, hallucination.
\end{IEEEkeywords}

\section{Introduction}
Text-to-speech (TTS) synthesis is a monotonic sequence-to-sequence task, maintaining a strict order between the input phoneme sequence and the output speech sequence. Moreover, the output speech sequence is at frame-level and one phoneme may correspond to multiple frames of speech, so the output sequence is significantly longer than its corresponding input phoneme sequence. Typical non-autoregressive neural text-to-speech models, such as FastSpeech 2 \cite{fastspeech2}, GradTTS \cite{gradtts}, UniCATS \cite{unicats} and NaturalSpeech 2 \cite{naturalspeech2}, integrate an explicit duration prediction module. Prior to training, the target duration is conventionally derived using the Viterbi forced alignment algorithm. 
% During training, this module is optimized by minimizing the distance between predicted and target durations. In the inference phase, the duration predictor module predicts the duration for each input phoneme, establishing the alignment between the input and output sequences accordingly. The encoded input phoneme sequence is then expanded to the frame level based on the predicted duration and is subsequently passed to the speech decoder. This mechanism enforces monotonic alignment constraints on the sequence-to-sequence process, ensuring robustness in the synthesis of speech.

Over the past two years, utilizing discrete speech tokens for speech generation is proposed in GSLM \cite{textlessnlp} and VQTTS \cite{vqtts}, paving the way for integrating cutting-edge language modeling techniques into TTS systems. Inspired by exceptional strides in natural language processing driven by decoder-only large Transformer models like GPT 3 \cite{gpt3} and the LLAMA 2 \cite{llama2}, Tortoise-TTS \cite{tortoise}, SPEAR-TTS \cite{speartts} and VALL-E \cite{valle} adopted the decoder-only architecture for TTS, achieving remarkable naturalness. SPEAR-TTS and VALL-E also have the ability to perform zero-shot speaker adaptation through in-context learning and auto-regressive (AR) continuation from a given speech prompt. Furthermore, these decoder-only TTS models, unlike the previous non-autoregressive neural TTS models, circumvent explicit duration modeling and the requirement for phoneme durations obtained before hand. This characteristic offers convenience and simplifies training process, especially when training on large scale datasets. However, the implicit duration modeling within these systems lacks the monotonic alignment constraints, often leading to hallucination issues like mispronunciation, word skipping and repeating.

% \begin{table}[t]
% \caption{Comparison between different types of TTS models.}
% \label{tab:merit_comparison}
% \begin{tabular}{c|ccc}
% \hline
%  & \textbf{Non-autoregressive TTS} &  \textbf{Decoder-only TTS} & \textbf{Ours} \\
%  & (e.g. FastSpeech 2)  & (e.g. VALL-E)  & (VALL-T) \\ \hline  \hline
% Free of explicit duration model &  $\times$ & $\checkmark$ & $\checkmark$ \\
% Monotonic alignment constraint   & $\checkmark$ & $\times$ & $\checkmark$ \\ \hline
% \end{tabular}
% \end{table}

% \begin{figure*}[t]
%   \centering
%   \subfigure[Model architecture.]{
%     \includegraphics[page=5,width=0.4\linewidth,trim=6cm 2.5cm 7cm 0.5cm,clip=true]{figure.pdf}
%     \label{transducer}
%   }
%   \subfigure[Alignment grid and a monotonic alignment path $\bar{\bm{y}}$ of length $T+U$.]{
%     \includegraphics[page=6,width=0.25\linewidth,trim=12cm 5.5cm 12cm 5cm,clip=true]{figure.pdf}
%     \label{grid}
%   }
%   \caption{The model architecture of previous modularized Transducer and its monotonic alignment path.}
%   \label{fig:transducer}
% \end{figure*}

In this work, we introduce VALL-T, a decoder-only generative Transducer model that combines the best of both worlds. It eliminates the need for obtaining phoneme duration before hand while still maintains the monotonic alignment constraint. 
Transducer \cite{transducer}, also known as RNN-T, is an existing training scheme designed specifically for monotonic sequence-to-sequence task. Typical Transducer model adopts a modularized architecture, composed of an encoder, a prediction network and a joint network, and has demonstrated success in automatic speech recognition (ASR) \cite{transducer_asr} as a classification task.
% However, such modularized architecture is less suited for TTS as a generation task. Further insights into this matter will be discussed in Chapter \ref{sec:vallt}.
Unlike existing modularized Transducers, VALL-T integrates all the three seperate modules into a single decoder-only Transformer architecture by introducing shifted relative positions, which is more suited for generation tasks. 
% Instead of slicing the hidden sequence from the encoder and prediction network at corresponding positions to construct the alignment grid, VALL-T uses additional relative position embeddings to indicate the positions of input phonemes. A relative position of 0 specifies the current phoneme being synthesized, enabling the construction of the alignment grid by iterating over all possible relative positions. 
During inference, VALL-T allows us to explicitly guide the monotonic generation process by shifting the relative positions from left to right and dramatically alleviate the hallucination issue in decoder-only TTS.

To the best of our knowledge, this is the first work that implements Transducer with a decoder-only Transformer architecture.
VALL-T presents several advantages compared to previous TTS models:
\begin{itemize}
    \item VALL-T introduces monotonic alignment constraints without altering the decoder-only architecture, leading to a better robustness against hallucination.
    \item VALL-T is capable of forced alignment given paired phoneme and speech sequences.
    \item The alignment between phoneme and speech sequences is aware during inference by tracking the timing of shifting the relative positions.
    % \item The alignment controllability of VALL-T during inference enables the utilization of untranscribed speech prompts within in-context learning framework.
\end{itemize}

\begin{figure*}[t]
\centering
\includegraphics[width=0.98\textwidth]{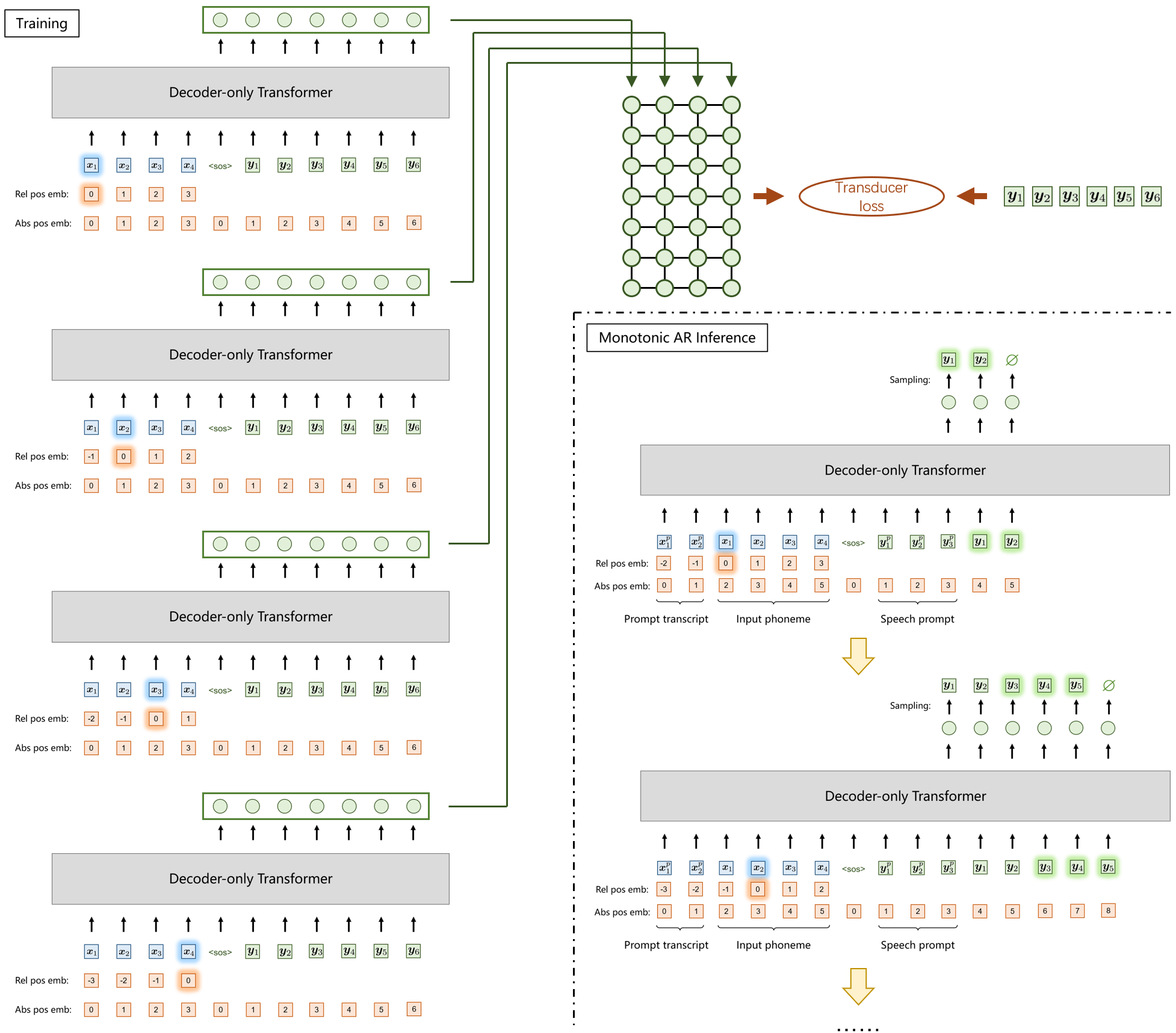}
\caption{The training and monotonic auto-regressive inference of VALL-T.}
\label{fig:vallt}
\end{figure*}

\section{Related work}

The Transducer model \cite{transducer}, also known as RNN-T, is designed for monotonic sequence-to-sequence tasks and comprises three components: an encoder, a prediction network, and a joint network. Here, the prediction network is an auto-regressive network, such as RNN and LSTM. Transducer model also introduces a special output token called blank, denoted as $\varnothing$, which signifies the alignment boundary between output and input sequence. We define $\mathcal{Y}$ as the vocabulary of output tokens and $\bar{\mathcal{Y}}=\mathcal{Y}\cup \{\varnothing\}$ as the extended vocabulary. Also, we denote the lengths of the input sequence $\bm{x}$ and output sequence $\bm{y}$ as $T$ and $U$ and the size of the extended vocabulary $\bar{\mathcal{Y}}$ as $\bar{V}$.

In the training phase, the encoder and prediction network encode the two sequences $\bm{x}$ and $\bm{y}$ respectively, yielding encoded hidden sequences $\bm{f}$ and $\bm{g}$. 
Then the joint network combines $\bm{f}$ and $\bm{g}$ at all possible positions to predict the corresponding next output token, constructing an alignment grid.
% $\bm{p}$ of shape $T\times (U+1)$ whose entry at $(t, u)$ is $\bm{p}_{t,u}$.
% Subsequently, the hidden vectors $\bm{f}_t$ and $\bm{g}_u$ at positions $t$ and $u$ respectively, then send them to the joint network to calculate the probability $\bm{p}_{t,u}=\mathbf{Pr}(\bar{\bm{y}}_{t+u} | \bm{f}_t, \bm{g}_u)$ for the next token prediction, where $\bar{\bm{y}}_{t+u}\in \bar{\mathcal{Y}}$. 
% % We combine all possible sliced hidden vectors of the two sequences, from $\bm{f}_0$ to $\bm{f}_{T-1}$ and from $\bm{g}_0$ to $\bm{g}_U$, constructing an alignment grid $\bm{p}$ of shape $T\times (U+1)$ whose entry at $(t, u)$ is $\bm{p}_{t,u}$.
% % % \footnote{The ``$+1$'' here comes from the additional \texttt{<sos>} at the beginning of the output sequence.}
Each path $\bar{\bm{y}}$ from the bottom left corner to the top right corner of the grid represents an alignment between $\bm{x}$ and $\bm{y}$, with a length of $T+U$. 
% % Figure \ref{grid} demonstrates an example of the alignment path where $\bar{\bm{y}}=[\bm{y}_1, \bm{y}_2, \varnothing, \bm{y}_3, \varnothing, \bm{y}_4, \bm{y}_5, \varnothing, \bm{y}_6, \varnothing]$.
The training criterion of Transducer model is to maximize the probability of $\mathbf{Pr}(\bm{y}|\bm{x})$, which is the summation of the probabilities of all possible alignment paths $\bar{\bm{y}}$. 
% Formally, $\mathbf{Pr}(\bm{y}|\bm{x}) = \sum_{\bar{\bm{y}}} \mathbf{Pr}(\bar{\bm{y}}|\bm{x}) = \sum_{\bar{\bm{y}}} \prod_{i=0}^{T+U-1} \mathbf{Pr}(\bar{\bm{y}}_{i} | \bm{f}_{t_i}, \bm{g}_{u_i})$.
% % , that is
% % \begin{equation}
% % \begin{aligned}
% % 	\mathbf{Pr}(\bm{y}|\bm{x}) & = \sum_{\bar{\bm{y}}} \mathbf{Pr}(\bar{\bm{y}}|\bm{x})  = \sum_{\bar{\bm{y}}} \prod_{i=0}^{T+U-1} \mathbf{Pr}(\bar{\bm{y}}_{i} | \bm{f}_{t_i}, \bm{g}_{u_i})
% % \label{eq:prob}
% % \end{aligned}
% % \end{equation}
% % where $\bm{f}_{t_i}$ and $\bm{g}_{u_i}$ are sliced hidden vectors at corresponding positions specified by the alignment path $\bar{\bm{y}}$. In practice, this probability can be efficiently calculated with dynamic programming.

In the inference phase, the prediction network auto-regressively predicts the next token, conditioning on the sliced input hidden vectors that slide from $\bm{f}_0$ to $\bm{f}_{T-1}$ whenever the blank token $\varnothing$ emerges.

\section{VALL-T: Decoder-Only Generative Transducer}
\label{sec:vallt}

Modularized Transducer model has demonstrated significant success in ASR \cite{transducer_asr}. Nevertheless, its suitability for generation tasks is limited. Typically, the joint network is a small network, comprising only one or a few linear projection layers, and the prediction network is LSTM or Transformer. This architecture introduces a limitation wherein the input condition is not incorporated into the generation process until it reaches the joint network. Worse still, the joint network is too small to effectively integrate input conditions into the generation process. Moreover, the modularized Transducer model utilizes slicing to denote specific positions. Consequently, the joint network is unable to explicitly perceive the input context, further making difficulties in achieving satisfactory performance for conditional generation tasks.

Therefore, VALL-T integrates the encoder, the prediction network and the joint network into one single decoder-only Transformer architecture and leverages relative position embedding to denote the corresponding positions. We discuss the training and inference details below.

% \definecolor{myorange}{RGB}{197,90,17}
\begin{figure*}[t]
  \centering
  \subfigure[Forward variable.]{
    \includegraphics[page=14,width=0.18\linewidth,height=6cm,trim=2.5cm 6cm 26cm 3cm,clip=true]{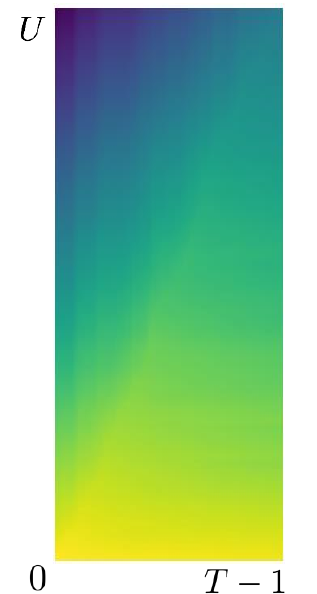}
    \label{alpha}
  }
  \subfigure[Backward variable.]{
    \includegraphics[page=14,width=0.18\linewidth,height=6cm,trim=17.1cm 6cm 11.4cm 3cm,clip=true]{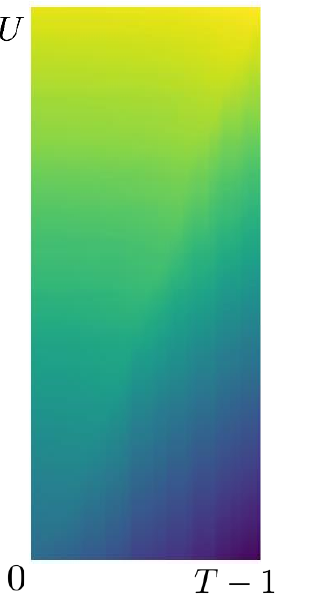}
    \label{beta}
  }
  \subfigure[Posterior probability.]{
    \includegraphics[page=15,width=0.18\linewidth,height=6cm,trim=2.5cm 6cm 26cm 3cm,clip=true]{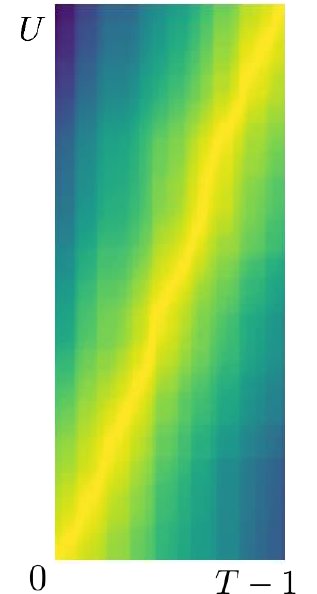}
    \label{prob}
  }
  \subfigure[Monotonic alignment path.]{
    \includegraphics[page=15,width=0.185\linewidth,height=6cm,trim=16.3cm 6cm 11.95cm 3cm,clip=true]{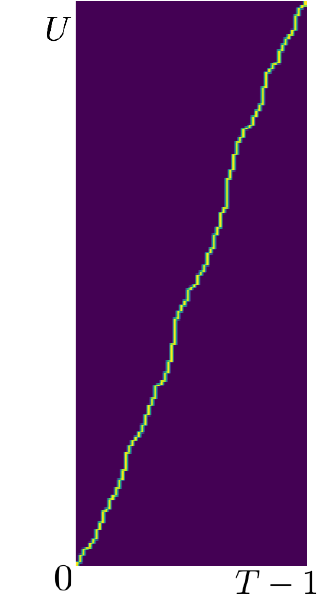}
    \label{path}
  }
  \caption{The plot of statistics for alignment analysis with VALL-T given both speech and its transcript.
  % In this example, the transcription is ``\texttt{His articulations usefully displaced and fashioned to bending the wrong way, \textcolor{myorange}{had received the education of a clown}, and could, like the hinges of a door, move backwards and forwards.}''.
  }
  \label{fig:alignment}
\end{figure*}

\subsection{Training}
\label{sec:training}

% Let ${\bm{x}}$  represent the input sequence and ${\bm{y}}$ represent the output sequence. Transducer \cite{transducer} introduces a special output token called blank, denoted as $\varnothing$, which signifies the alignment boundary between output sequence ${\bm{y}}$ and input sequence $\bm{x}$. We define $\mathcal{Y}$ as the vocabulary of output tokens and $\bar{\mathcal{Y}}=\mathcal{Y}\cup \{\varnothing\}$ as the extended vocabulary. Also, we denote the lengths of the input and output sequence as $T$ and $U$, and the size of the extended vocabulary $\bar{\mathcal{Y}}$ as $\bar{V}$.

We use a decoder-only architecture for VALL-T. Similar to the approach in the previous work VALL-E, we concatenate the input phoneme and output speech tokens along the time axis and present them to the model as a unified sequence. Unlike traditional RNN and LSTM architectures, the Transformer lacks a specific time order for input tokens, relying instead on position embeddings to indicate their positions. The position indices for the input sequence range from $0$ to $T-1$ and are converted into position embeddings through a sinusoidal function \cite{transformer}. Similarly, the output sequence adopts position indices from $0$ to $U$, including an additional \texttt{<sos>} token at the beginning. Following VALL-E, we utilize a triangular attention mask for the output sequence, facilitating auto-regressive generation. This mask ensures that each speech token attends to only previously generated tokens, maintaining a proper sequential order in the output.

Beyond the typical absolute position indices starting from 0, we introduce additional relative position indices in VALL-T for input tokens. The relative position index $0$ specifies the current phoneme under synthesis. The phonemes to its left are assigned negative position indices starting from $-1$, while those to its right are assigned positive position indices starting from $1$. These relative position indices are converted to relative position embeddings with a same sinusoidal function as the absolute position indices. The resulting absolute and relative position embeddings are added to the input phoneme embeddings and subsequently presented to the decoder-only Transformer. By adopting this approach, the model gains awareness of the phoneme presently undergoing synthesis, specifically the one assigned a relative position of $0$, and the phonemes serving as its preceding and subsequent contexts.

The blank token $\varnothing$ serves as a marker denoting the end of each phoneme's generation. Consequently, the output projection following the decoder-only Transformer projects the hidden sequence into the extended vocabulary size of $\bar{V}$. The projected hidden sequence, with a length of $U+1$, undergoes a Softmax function to yield a sequence representing the output distribution.

Transducer loss is computed on top of an alignment grid in the shape of $T \times (U+1)$.
Different from the method for constructing the alignment grid in the modularized Transducer, VALL-T treats the output sequence of the decoder-only model as a column within this grid, corresponding to the phoneme indicated by the relative position index 0. Through iterating over all possible relative positions, as illustrated in Figure \ref{fig:vallt}, the alignment grid is constructed. This enables us to compute the Transducer loss on top of the grid to maximize the probability of $p(\bm{y}|\bm{x})$.

% Illustrated in Figure \ref{fig:vallt}, we iteratively assign relative position $0$ to each of the $T$ phonemes and subsequently stack every output sequence, each of length $U+1$. This stacking process results in a matrix $\bm{p}$ of shape $T\times (U+1)$. The optimization of VALL-T utilizes the Transducer loss, calculated using this matrix and the ground-truth speech tokens, 

\begin{table*}[h]
\caption{The performance of zero-shot TTS.}
\label{tab:tts}
\begin{adjustbox}{width=1.6\columnwidth,center}
\begin{tabular}{c|ccccc}
\hline
\textbf{Method}  & \textbf{WER(\%)} $\downarrow$  & \textbf{MCD} $\downarrow$  & \textbf{Naturalness MOS} $\uparrow$ & \textbf{Similarity MOS} $\uparrow$  & \textbf{SECS} $\uparrow$ \\ \hline  \hline
 Ground-truth        & 1.92  &  0   &  4.63 $\pm$ 0.07  & 4.23 $\pm$ 0.10  &  0.837 \\
Encodec resynthesis  & 2.08  & 2.50 &  4.55 $\pm$ 0.07  & 4.19 $\pm$ 0.11  &  0.835 \\
NAR resynthesis      & 3.75  & 2.95 &  4.44 $\pm$ 0.07  & 4.24 $\pm$ 0.10  &  0.846 \\ \hline
Transduce and Speak  & 6.14  & 4.38 &  4.07 $\pm$ 0.10  & 4.02 $\pm$ 0.11  &  0.838 \\
VALL-E               & 5.80  & 4.00 &  4.25 $\pm$ 0.08  & 4.12 $\pm$ 0.10  &  \textbf{0.857} \\
VALL-T (ours)        & \textbf{4.16} & \textbf{3.98} & \textbf{4.26 $\pm$ 0.08} &  \textbf{4.21 $\pm$ 0.09} & 0.849 \\ \hline
\end{tabular}
\end{adjustbox}
\end{table*}

\subsection{Monotonic auto-regressive inference}
\label{sec:inference}
Let us first consider the auto-regressive inference process without a speech prompt. Initially, the relative position $0$ is designated to the first phoneme, starting the speech generation from the \texttt{<sos>} token. The model then auto-regressively produces speech tokens based on the input phoneme tokens and previously generated speech tokens until the blank token $\varnothing$ emerges. The emergence of $\varnothing$ denotes the completion of the first phoneme's generation and triggers a shift in relative positions. We iteratively conduct the above process until the appearance of $\varnothing$ for the last phoneme, indicating the conclusion of the entire generation process for the input phoneme sequence. 
Since the model is encouraged to generate speech tokens for the phoneme assigned relative position $0$ by Transducer loss during training, the step-by-step shifting operation during decoding facilitates the monotonic generation process and consequently enhance the robustness against hallucination.

Next, we consider the integration of the speech prompt for zero-shot speaker adaptation. Following the approach used in VALL-E, the phoneme transcript of the speech prompt is placed at the start of the input sequence, while the speech prompt itself is positioned at the beginning of the output sequence. The two sequences are followed by the input phonemes to be generated and their corresponding output speech tokens respectively. Given that the speech prompt are provided, we assign the relative position 0 to the first phoneme right after the prompt transcript, as shown in Figure \ref{fig:vallt}, and perform speech continuation. Likewise, the relative positions undergo a shift each time $\varnothing$ emerges, repeating until the generation for the final phoneme is completed.

\section{Experiments and Results}

\subsection{Setup}

VALL-T is compatible with any speech tokenizers. In our experiments, we use the Encodec \cite{encodec} speech tokenizer, following VALL-E, to ensure a fair comparison between our model and VALL-E and eliminate other factors that could potentially affect our analysis. The speech tokens are extracted in 50Hz and the sampling rate of output waveforms is 16k. It comprises 8 residual vector quantization (RVQ) indices for each frame. We also follow the approach introduced in VALL-E that predicts the sequence of the first RVQ index with the auto-regressive models and then predicts the remaining 7 RVQ indices conditioned on the first RVQ index with a separate non-auto-regressive (NAR) model. Both the input and output sequences are encoded with BPE \cite{bpe} algorithm to shorten sequence lengths and diminish GPU memory consumption. VALL-T adopts an identical architecture to VALL-E, containing 12 layers of Transformer blocks. Each block comprises 12 attention heads and has a hidden dimension of 1024.

We use LibriTTS \cite{libritts} dataset in our experiments, which is a multi-speaker transcribed English speech dataset. Its training set consists of approximately 580 hours of speech data from 2,306 speakers. We train our model for 40 epochs using a ScaledAdam \cite{zipformer} optimizer. The learning rate scheduler is Eden \cite{zipformer} with a base learning rate of $0.05$, an epoch scheduling factor of 4 and a step scheduling factor of 5000. We use 8 A800 GPUs for the training.

% \begin{table*}[h]
% \caption{Zero-shot TTS with untranscribed speech prompt.}
% \label{tab:untranscribed}
% \begin{adjustbox}{width=1.6\columnwidth,center}
% \begin{tabular}{cc|ccccc}
% \hline
% \multirow{2}{*}{\textbf{Method}} & \textbf{Pseudo Prompt} & \multirow{2}{*}{\textbf{WER(\%)} $\downarrow$}  & \multirow{2}{*}{\textbf{MCD} $\downarrow$}  & \multirow{2}{*}{\textbf{Naturalness MOS} $\uparrow$} & \multirow{2}{*}{\textbf{Similarity MOS} $\uparrow$}    & \multirow{2}{*}{\textbf{SECS} $\uparrow$} \\ 
%      &  \textbf{Transcription}  &    &  &   &    &  \\ \hline  \hline
% % Ground-truth & -  & 1.92  &   &    &  0  & 0.837 \\ \hline
% \multirow{2}{*}{VALL-E} & $\times$ & 68.22 & 4.97 & - & -  & 0.795 \\
%                         & $\surd$  & 21.01 & 4.28 & - & -  & 0.836 \\ \hline
% \multirow{2}{*}{VALL-T} & $\times$ & 30.86 & 4.43 & - & -  & 0.836\\
%                         & $\surd$  & \textbf{3.48}  &  \textbf{3.97} & 4.29 $\pm$ 0.09 & 4.14 $\pm$ 0.10 & \textbf{0.848} \\ \hline
% \end{tabular}
% \end{adjustbox}
% \end{table*}

\subsection{Alignment analysis}
\label{sec:alignment_anal}

We first do alignment analysis to check if relative position embedding in VALL-T indicates the alignment as expected. Given the speech $\bm{y}$ and its transcript $\bm{x}$, we iterate over all relative positions and calculate the alignment grid $\bm{p}$ of output distributions in the shape of $T\times (U+1)$. Then we calculate the forward variables, backward variables and posterior probabilities accordingly. The definitions of forward variable, backward variables, and posterior probabilities have been introduced in \cite{transducer}. 
% We describe our calculation for these values in Appendix \ref{sec:calculate}.

In Figure \ref{fig:alignment}, we illustrate an example of the forward variable, backward variable, and posterior probability for VALL-T, with darker colors indicating lower values. The values are plotted on a logarithmic scale. In Figure \ref{alpha} and \ref{beta}, we can see a faint bright line on the diagonal of the two graphs.
%Analyzing the forward variable map in Figure \ref{alpha}, we observe that the point $(0, 0)$ is the lightest, and the colors darken rapidly along the vertical axis. This pattern suggests that the model is unlikely to produce the speech tokens without shifting the relative positions. Along the horizontal axis, the color doesn't darken as fast, primarily because the model tends to assign a relatively high probability to the blank token, given its frequent occurrence. Importantly, a discernible light line emerges along the diagonal direction, indicating that achieving points along this line is more probable than their surrounding points.
%Similarly, in Figure \ref{beta} of the backward variable map, the top-right corner appears the lightest, and a noticeable light line along the diagonal direction is also evident.

Pixel-wise summing the values from Figure \ref{alpha} and Figure \ref{beta} produces Figure \ref{prob}, which represents the posterior probability. The diagonal line becomes much clearer in this composite figure, indicating that VALL-T correctly models the alignment between the input and output sequences with relative position embeddings. Accordingly, VALL-T is capable of forced alignment, where the most probable path from the bottom-left corner to the top-right corner in the posterior probability map serves as the alignment path. The alignment path for this example is depicted in Figure \ref{path}.
% where the waveform segment highlighted in orange aligns with the corresponding transcription in the same orange color. 
Since ground-truth labels for alignment are unavailable, our alignment analysis here only focuses on qualitative aspects.

%During inference, acquiring alignment is straightforward by keeping track of the shifting history of relative positions at each decoding time step. In Figure \ref{fig:align_aware_dec}, we illustrate an example of the alignment path during decoding, revealing that the path basically follows a diagonal trajectory. This alignment-aware decoding capability empowers us to leverage untranscribed speech prompts and facilitates streaming generation, a topic we will delve into later.
%
%
%\begin{figure}[ht]
%\centering
%\includegraphics[page=16,width=0.65\linewidth,trim=15cm 6cm 9cm 2.9cm,clip=true]{figure.pdf}
%\caption{Alignment-aware decoding given transcript. In this example, the transcript is ``\texttt{He took no notice of her; he looked at me, but 
%as if, instead of me, he saw what he spoke of.}''}
%\label{fig:align_aware_dec}
%\end{figure}

\subsection{Evaluation on zero-shot TTS}
\label{sec:zeroshottts}

In this section, we conduct an evaluation of our models on zero-shot TTS task. The task refers to synthesizing speech in the voices of unseen speakers given speech prompts and their corresponding transcripts. Our test set uses a same test set as in \cite{unicats}, containing 500 utterances and involving 37 speakers from the LibriTTS test set. Each speaker is assigned a specific speech prompt.
Before assessing the performance of our models, we conduct speech resynthesis using our Encodec to evaluate the speech tokenizer. We also do an experiment named ``NAR resynthesis''. In this experiment, we send the ground-truth first RVQ index to the NAR model for predicting the remaining 7 RVQ indices. Then, we convert all the 8 RVQ indices to waveform using the Encodec decoder. The purpose of the NAR resynthesis experiment is to demonstrate the performance degradation introduced by the NAR model, so we can better analyze the results of the entire pipelines, where the AR models are the primary focus of our paper.

The baselines of this experiment include two models. One is the popular decoder-only TTS model VALL-E and another is the recently proposed TTS model with a modularized Transducer achitecture called ``Transduce and Speak'' \cite{transduceandspeak}. The main evaluation metric in this paper is the word error rate (WER). In our evaluation process, we first synthesize speech for the test set, and then perform speech recognition using a well-known ASR model, Whisper\footnote{https://huggingface.co/openai/whisper-medium} \cite{whisper}. The transcripts obtained from the ASR model are then compared to the ground-truth input text to calculate the word error rate. 
Table \ref{tab:tts} shows that VALL-T attains significant lower WER than baselines, which is a 28.3\% relative reduction when compared to VALL-E and is only 0.41 higher than NAR resynthesis, suggesting the robustness of VALL-T. 

Additionally, we present the mel-cepstral distortion (MCD), Speaker Embedding Cosine Similarity (SECS) and Mean Opinion Score (MOS) for naturalness and speaker similarity. Note that the SECS here is evaluated between the generated speech and the provided speech prompt, not the corresponding ground-truth speech, since our goal is to emulate solely the voice of the given prompt. It is obtained using a pretrained speaker verification model\footnote{https://github.com/resemble-ai/Resemblyzer}. In MOS tests, 15 listeners were tasked with rating each utterance on a scale from 1 to 5, with higher scores indicating better naturalness and similarity. We can observe that VALL-T performs comparably to or even better than the baselines on most of these metrics when WER dramatically decreases.

\section{Conclusion}

In this research, we present VALL-T, a decoder-only generative Transducer model designed to improve the robustness and controllability of TTS models. VALL-T incorporates monotonic alignment constraints into the decoder-only TTS framework, enabling implicit modeling of phoneme durations. Threfore, this model eliminates the need for acquiring phoneme durations before training. VALL-T supports forced alignment given input phonemes and the corresponding output speech by searching the best path on the posterior probability map.
% This alignment is controllable during inference, facilitating zero-shot synthesis with untranscribed speech prompts.
% Additionally, VALL-T exhibits the capability of streaming generation, coupled with an aligned context window for synthesizing lengthy speech. 
% These features make VALL-T a powerful model for TTS applications.

% Despite all the above advantages, VALL-T still faces a drawback: its training cost is higher compared to VALL-E. This is primarily due to the need to iterate over all possible positions. The increase is proportional to the length of the phoneme sequence. 

% There are many potential ways of reducing the training cost of Transducer in the literature, such as pruning techniques to eliminate impossible alignment paths \cite{transducer_pruning} and multi-stage training strategies \cite{transducer_multistage}. We will explore the applications of these techniques in VALL-T in the future works.

\section*{Acknowledgment}

This work was supported by the China NSFC Project (No. 92370206), the Key Lab of Suzhou on Linguistic Computing (SZS2024005), Shanghai Municipal Science and Technology Major Project (2021SHZDZX0102) and Open Project of the Key Laboratory of Artificial Intelligence,
Ministry of Education (AI202405).

\bibliographystyle{IEEEtran}
\bibliography{main}

% \vspace{12pt}
% \color{red}
% IEEE conference templates contain guidance text for composing and formatting conference papers. Please ensure that all template text is removed from your conference paper prior to submission to the conference. Failure to remove the template text from your paper may result in your paper not being published.

\end{document}